# Anisotropic upper critical field in the van der Waals superconducting quasicrystal $(Ta_{0.7}Nb_{0.3})_{1.6}Te$

Koki Kasai[1], Yuki Tokumoto[1, *], Taichi Terashima[2], Takako Konoike[2], and Keiichi Edagawa[1]

[1]Institute of Industrial Science, The University of Tokyo, Tokyo 153-8505, Japan
[2]Research Center for Materials Nanoarchitectonics (MANA), National Institute for Materials Science, Tsukuba 305-0003, Japan

**ABSTRACT**. We investigated the upper critical field  of a large single grain of the Nb-substituted van der Waals layered quasicrystal $(Ta_{0.7}Nb_{0.3})_{1.6}Te$. The sample exhibits a sharp superconducting transition at $T_c$ = 1.35 K, the highest value reported to date among quasicrystal superconductors. The angular dependence of the critical field exhibits a pronounced criterion dependence: the field determined using the 10% $R_N$ ($R_N$: normal-state resistance) criterion is well described by the anisotropic Ginzburg–Landau model, whereas those determined using the 65% and 90% $R_N$ criteria exhibit Tinkham-like angular dependence characteristic of two-dimensional superconductivity. The high-field part of the resistive transition is well described by a surface-superconductivity model and exhibits a pronounced excitation-current dependence for magnetic fields close to the *ab* plane, supporting the presence of surface superconductivity on the quasiperiodic *ab*-plane surfaces. The bulk $H_{c2}$ is strongly anisotropic, with the in-plane $H_{c2}$ exceeding the weak-coupling Pauli limit by a factor of approximately 2.5. For both field orientations, $H_{c2}(T)$ deviates upward from the conventional dirty-limit Werthamer–Helfand–Hohenberg prediction at low temperatures. A phenomenologically modified Ginzburg–Landau–Abrikosov–Gorkov model incorporating a spatial distribution of the electronic diffusivity substantially improves the description of $H_{c2}(T)$, suggesting that spatial variations in electronic transport properties may contribute to its anomalous temperature dependence.

## I. INTRODUCTION.

Quasicrystals (QCs) are a class of materials that possess long-range order but lack translational periodicity [1]. Because Bloch's theorem, which assumes translational periodicity of the atomic lattice, cannot be directly applied to QCs, the Bloch wave vector is no longer a good quantum number. Since the discovery of QCs, an important question has been whether long-range electronically ordered states, such as superconductivity, can exist in such aperiodic systems. In 2018, bulk superconductivity was discovered for the first time in the Al–Zn–Mg icosahedral QC [2].

The nature of superconductivity in QCs has attracted considerable interest and extensive theoretical studies have been carried out in recent years. A variety of unconventional phenomena have been predicted, including Cooper pairing beyond the conventional ($\mathbf{k}\uparrow$, $-\mathbf{k}\downarrow$) pairing scheme [3], a novel superconducting state in which the superconducting order parameter exhibits a fractal-like spatial sign modulation at low temperatures under magnetic fields [4], superconducting states with spatially inhomogeneous order parameters [5], intrinsic vortex pinning [6], Cooper pairs with combinations of orbital and spin components that cannot be realized in periodic crystals [7], and topological superconductivity [8].

In 2024, we discovered superconductivity in a Ta–Te van der Waals layered dodecagonal QC [9]. Unlike the Al–Zn–Mg QC, which is thermodynamically metastable and has an extremely low superconducting transition temperature ($T_c$) of approximately 50 mK, the Ta–Te QC is thermodynamically stable and has a $T_c$ of approximately 1 K. It therefore provides an attractive platform for investigating the superconductivity in QCs.

For polycrystalline Ta–Te QC samples, we subsequently observed an anomalous linear increase of the upper critical field down to 0.04 K [10]. The anomalous superconducting properties of Ta–Te QCs have also stimulated recent theoretical studies on spin–orbit effects in quasicrystalline superconductors [11]. However, measurements on polycrystalline samples average over randomly oriented grains, making it difficult to determine the intrinsic anisotropy of the superconducting state.

More recently, the superconducting Ta–Te QC family has been expanded to various ternary and quaternary compounds through third-element substitution, considerably broadening the compositional space available for investigating superconductivity in van der Waals layered QCs [12]. Among the various third-element substitutions, Nb substitution was found to systematically enhance $T_c$ with increasing Nb content.

To clarify the intrinsic anisotropy of the upper critical field without orientational averaging and to gain further insight into its anomalous temperature dependence, measurements on a sufficiently large single grain are required. Here, we report the preparation of a sufficiently large single grain of the Ta–Nb–Te QC, $(Ta_{0.7}Nb_{0.3})_{1.6}Te$, which exhibits a superconducting transition at $T_c$ =1.35 K, the highest $T_c$ reported to date among QC superconductors. We use this single grain to investigate the temperature and angular dependences of its upper critical field.

## II. METHODS

The $(Ta_{0.7}Nb_{0.3})_{1.6}Te$ samples were synthesized by reaction sintering. A mixture of $TaTe_2$, Nb, Ta, and iodine (used as a reaction promoter) was pressed into a pellet and heat-treated. High-temperature sintering was carried out in a Mo crucible sealed by arc melting under an Ar atmosphere, using a two-step heat-treatment process: first at 1100 °C for 120 h, followed by 1400 °C for 120 h. Phase identification was performed by powder X-ray diffractometry (XRD).

For the electrical resistance measurements, a large single grain grown on the inner wall of the Mo crucible was extracted. Electrical contacts were made with silver conducting paste, and the electrical resistance was measured using the standard four-probe method.

The temperature and magnetic-field dependences of the electrical resistance were measured down to 0.04 K using a dilution refrigerator equipped with a 20-T superconducting magnet. To investigate the magnetic-field-angle dependence of the upper critical field, the sample was mounted on a top-loading probe and rotated in situ on a rotation platform. The electrical current was always applied perpendicular to the magnetic field.

## III. RESULTS

After heat treatment, single-crystalline grains of the $(Ta_{0.7}\ Nb_{0.3})_{1.6}Te$ sample with an in-plane size of approximately 1 mm and a thickness of several hundred micrometers were found to have grown on the inner wall of the Mo crucible, separately from the sintered pellet (see Supplemental Material [13]). Figure 1 shows the experimental powder XRD pattern of the sample together with the calculated profile for the dodecagonal QC phase. The calculated profile was obtained from that of the $Ta_{97}Te_{60}$ approximant using the quasicrystal–approximant relationship, with the effects of phason disorder and stacking disorder along the *c*-axis taken into account. Details of the calculation are described in the Supplemental Material [13]. The calculated profile reproduces the experimental pattern reasonably well. In particular, in addition to the major diffraction peaks, the broad intensity distribution observed over $2\theta \simeq 30^\circ$–$50^\circ$ is also well reproduced. The calculation shows that this broad feature arises from the superposition of a large number of weak diffraction peaks. All discernible diffraction peaks can be indexed to the dodecagonal QC phase, with no detectable reflections attributable to secondary phases, indicating the formation of a single-phase dodecagonal QC sample.

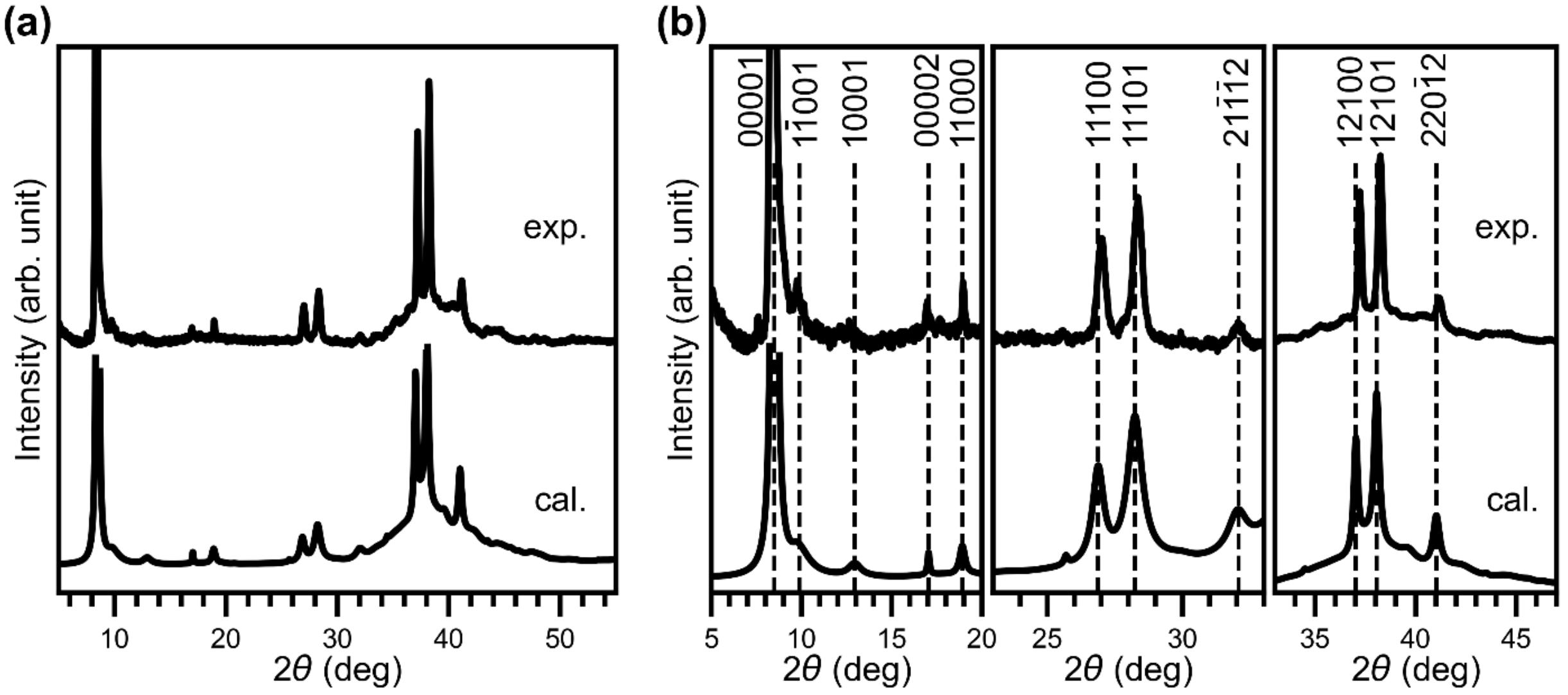

FIG 1. (a) Experimentally obtained powder XRD profile of the synthesized $(Ta_{0.7}Nb_{0.3})_{1.6}Te$ sample and the calculated profile for the dodecagonal QC phase. (b) Magnified view of the data shown in (a).

The single grain used for the measurements had an in-plane size of approximately 1 mm in the quasiperiodic (*ab*) plane and a thickness of approximately 350 μm (inset in Fig. 2(a)). Figure 2(a) shows the temperature dependence of the electrical resistance $R$ from 293 K down to 37 mK. The resistance gradually increases with decreasing temperature and abruptly drops to zero at low temperature. This behavior is similar to that observed in polycrystalline bulk Ta–Te-based binary, ternary, and quaternary QCs [9,12]. Figure 2(b) shows an enlarged view of the temperature dependence of $R$ between 1.1 and 1.7 K. At $T$ = 1.5 K, $R$ = 24 mΩ, corresponding to a resistivity $\rho$ of approximately 1 mΩ cm. The resistance drops sharply and reaches zero below 1.35 K, indicating a superconducting transition. The 10–90% transition width is as narrow as 0.02 K. The $T_c$ of 1.35 K is the highest reported so far among Ta–Te-based QC superconductors.

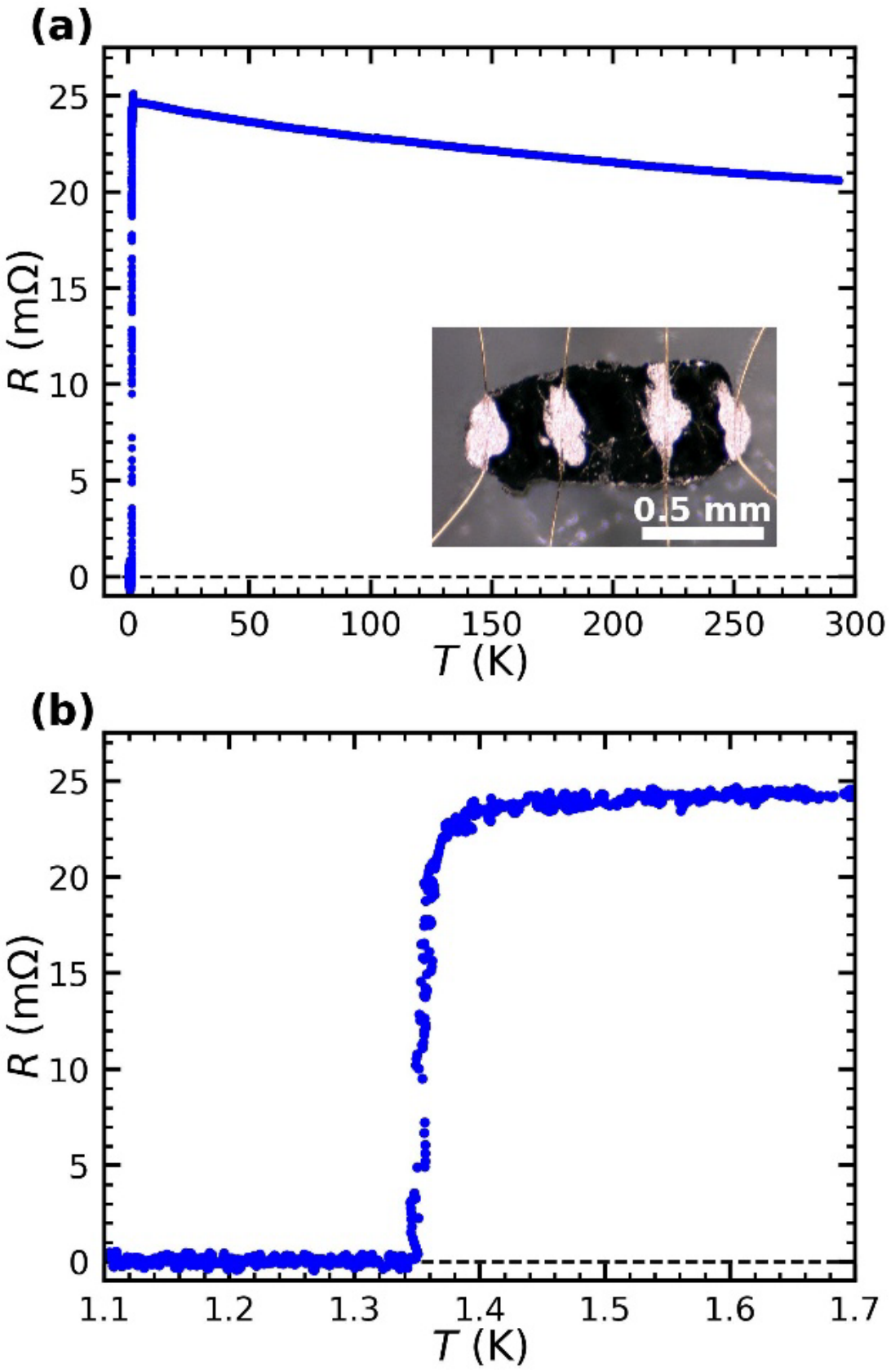

FIG 2. (a) Temperature dependence of the electrical resistance $R$ from 37 mK to 293 K. The inset shows the single-grain sample with four electrical contacts. (b) Enlarged view of the superconducting transition between 1.1 and 1.7 K.

Magnetoresistance measurements were performed at various temperatures and at different angles ($\theta$) between the applied magnetic field $H$ and the normal to the quasiperiodic (*ab*) planes. Figure 3(a) shows the magnetic-field dependence of the resistance measured at various angles $\theta$ at an average temperature of 45 mK. During the series of measurements at different field angles, the sample temperature varied between 34 and 51 mK. However, as will be shown later, the critical field exhibits only weak temperature dependence below 50 mK. Therefore, this temperature variation has a negligible influence on the angular dependence. To avoid self-heating of the sample, the measurement current was reduced to 5 μA, corresponding to a current density of approximately $3\times10^{-3}$ A cm$^{-2}$. For magnetic fields applied parallel to the *ab* planes (in-plane fields, $\theta$ = 90°), the superconducting transition is considerably broadened. In contrast, for magnetic fields close to the out-of-plane direction ($\theta \approx 0°$), a two-step superconducting transition is observed. For each *R*-*H* curve, the magnetic fields corresponding to 10%, 65%, and 90% of the normal-state resistance $R_N$ were determined and are plotted in the upper three panels of Fig. 3(b). For all three criteria, the critical field reaches its maximum when the magnetic field is parallel to the *ab* planes, corresponding to $\theta$=90°. The angular dependence of the critical field defined by the 10% $R_N$ criterion is well described by the anisotropic Ginzburg–Landau (GL) model [14]:

$$H_{c2}(\theta) = \frac{H_{c2}^{\perp}}{\sqrt{\cos^2\theta + \frac{\sin^2\theta}{\gamma^2}}}, \tag{1}$$

where $H_{c2}^{\perp}$ is the upper critical field for an out-of-plane field, and $\gamma = H_{c2}^{\parallel}/H_{c2}^{\perp}$ is the anisotropy parameter, with $H_{c2}^{\parallel}$ denoting the upper critical field for an in-plane field. In contrast, the angular dependences of the critical fields defined by the 65% and 90% $R_N$ criteria deviate from the anisotropic GL model and are better described by the Tinkham model for two-dimensional superconductors [15]:

$$\left(\frac{H_{c2}(\theta)\sin\theta}{H_{c2}^{\parallel}}\right)^2 + \left|\frac{H_{c2}(\theta)\cos\theta}{H_{c2}^{\perp}}\right| = 1. \tag{2}$$

Because a Tinkham-like angular dependence can also arise from surface superconductivity, the high-field part of the resistive transition was further compared with the angular dependence of the surface superconducting critical field $H_{c3}$ [16], assuming surface superconductivity on the *ab*-plane surfaces. Rewriting the expression of Yamafuji et al. [16] in terms of the angle $\theta$ defined above, we obtain

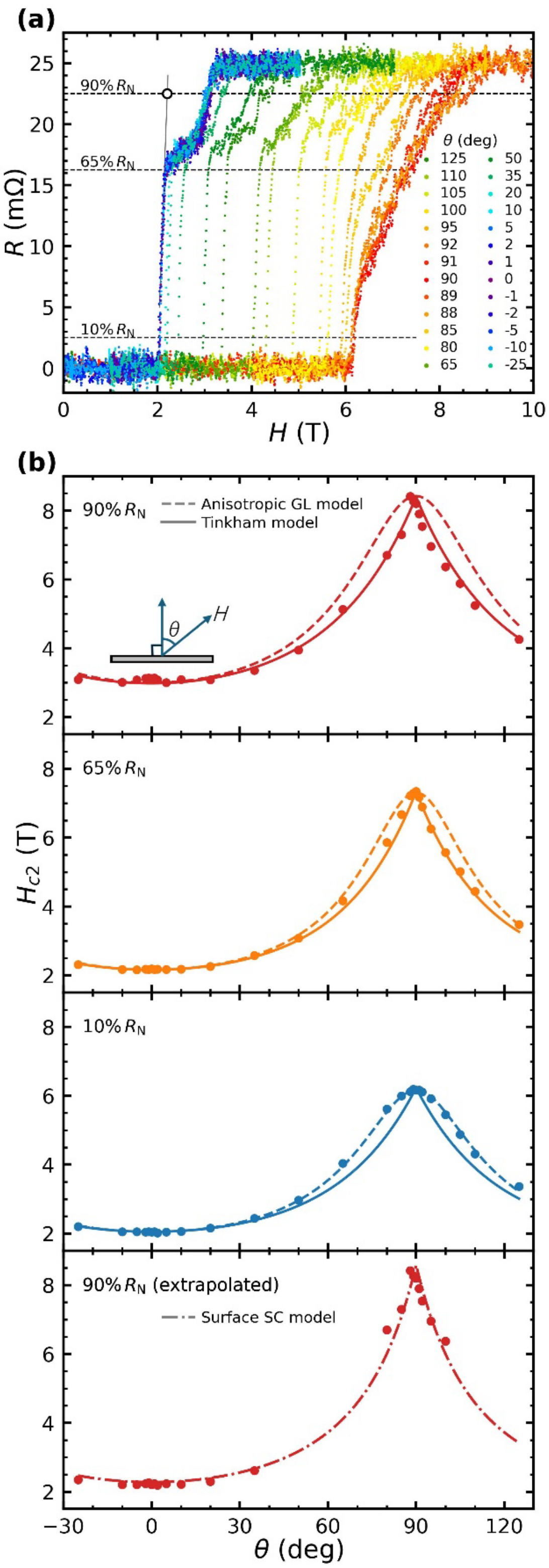

FIG 3. (a) Magnetic-field dependence of the resistance $R$ at various angles $\theta$ at an average temperature of 45 mK. (b) Angular dependence of the critical fields. The inset schematically illustrates the definition of $\theta$. The upper three panels show the fields determined using the 90%, 65%, and 10% $R_N$ criteria, respectively. Dashed and solid curves represent the anisotropic GL and Tinkham models, respectively. The bottom panel shows an alternative set of 90% $R_N$ fields obtained by extrapolating the lower-field transition near $H\perp ab$. The dash-dotted curve represents the surface-superconductivity model.

$$\left|\frac{H_{c3}(\theta)\sin\theta}{H_{c3}^{\parallel}}\right|^2 [1 + |\cot\theta|(1 - |\cos\theta|)] + \left|\frac{H_{c3}(\theta)\cos\theta}{H_{c2}^{\perp}}\right| = 1. \quad (3)$$

For -25° ≤ $\theta$ ≤ 35°, where a two-step transition was observed, an alternative 90% $R_N$ field was estimated by linearly fitting the lower-field transition between 20% and 60% $R_N$ and extrapolating it to the 90% $R_N$ level. For 75° ≤ $\theta$ ≤ 95°, the 90% $R_N$ fields were determined directly from the measured curves. No values were assigned at the remaining angles. For clarity, the extrapolation is illustrated only for $\theta$ = 0° in Fig. 3(a). The angular dependence obtained in this way, shown in the bottom panel of Fig. 3(b), is well described by the surface-superconductivity model. To further examine this interpretation, we investigated the excitation-current dependence of the resistive transition.

Figure 4 shows the magnetic-field dependence of the resistance measured with excitation currents of 5, 10, and 50 μA for in-plane and out-of-plane fields. For in-plane fields, the broad resistive transition exhibits a pronounced current dependence: the intermediate- and high-resistance parts of the transition shift appreciably with increasing excitation current, whereas the low-resistance part is much less affected. In contrast, for out-of-plane fields, the characteristic two-step transition shows a much weaker current dependence. These qualitatively different current dependences suggest that the broad transition for in-plane fields and the two-step transition for out-of-plane fields have different origins.

Figures 5(a) and 5(b) show $R$–$H$ curves measured at various temperatures for in-plane and out-of-plane fields, respectively. The superconducting transition shifts to lower fields with increasing temperature. Figure 5(c) shows the temperature dependences of $H_{c2}^{\parallel}$ and $H_{c2}^{\perp}$ determined using the 10% $R_N$ criterion, which are considered to predominantly reflect bulk superconductivity. For both field orientations, $H_{c2}$ exhibits only weak temperature dependence below 0.1 K. The zero-temperature upper critical fields are estimated to be $H_{c2}^{\parallel}(0) = 6.3$ T and $H_{c2}^{\perp}(0) = 2.0$ T. Using these values, the coherence lengths parallel and perpendicular to the $ab$ planes were estimated to be $\xi_{ab}$ = 13 nm and $\xi_c$ = 4.1 nm, respectively, from the relations $H_{c2}^{\parallel} = \frac{\Phi_0}{2\pi\xi_{ab}\xi_c}$ and $H_{c2}^{\perp} = \frac{\Phi_0}{2\pi\xi_{ab}{}^2}$, where $\Phi_0$ is the quantized magnetic flux. This corresponds to an anisotropy parameter of $\gamma = 3.2$. While $H_{c2}^{\perp}(0)$ remains below the Pauli limit, $H_P(0) = 2.51$ T, $H_{c2}^{\parallel}(0)$ exceeds it by a factor of approximately 2.5. The temperature dependences of $H_{c2}^{\parallel}$ and $H_{c2}^{\perp}$ were fitted using the Werthamer–Helfand–Hohenberg (WHH) theory [17]. According to the WHH theory, which takes spin paramagnetism and spin–orbit scattering into account, $H_{c2}$ in the dirty limit (mean free path $l \ll$ coherence length $\xi_0$) is described by the following equation:

$$\ln\frac{1}{t} = \left(\frac{1}{2} + \frac{i\lambda_{SO}}{4\gamma}\right)\psi\left(\frac{1}{2} + \frac{\bar{h} + \frac{1}{2}\lambda_{SO} + i\gamma}{2t}\right) + \left(\frac{1}{2} - \frac{i\lambda_{SO}}{4\gamma}\right)\psi\left(\frac{1}{2} + \frac{\bar{h} + \frac{1}{2}\lambda_{SO} - i\gamma}{2t}\right) - \psi\left(\frac{1}{2}\right), \quad (4)$$

where $t = T/T_c$, $\psi$ is the digamma function, $\gamma \equiv \left[(\alpha\bar{h})^2 - \left(\frac{1}{2}\lambda_{SO}\right)^2\right]^{\frac{1}{2}}$, and $\bar{h} = \frac{4}{\pi^2}h^* = \frac{4H_{c2}}{\pi^2(-dH_{c2}/dt)_{t=1}}$. The parameters $\lambda_{SO}$ and $\alpha$ represent the effects of spin–orbit scattering and spin paramagnetism, respectively. Within the WHH framework, the best agreement with the experimental data was obtained for $\alpha$ = 0, indicating that Pauli paramagnetic pair breaking is negligible within this description. For both in-plane and out-of-plane fields, the experimental $H_{c2}$ deviates upward from the WHH prediction below approximately 0.8 K. These results suggest that $H_{c2}$ cannot be fully described within the conventional dirty-limit WHH framework. As shown in Fig. 5(c), the low-temperature $H_{c2}(T)$ data are better reproduced by the phenomenologically modified Ginzburg–Landau–Abrikosov–Gorkov (GLAG) model proposed by Carter et al. [18], as discussed in detail in the following section.

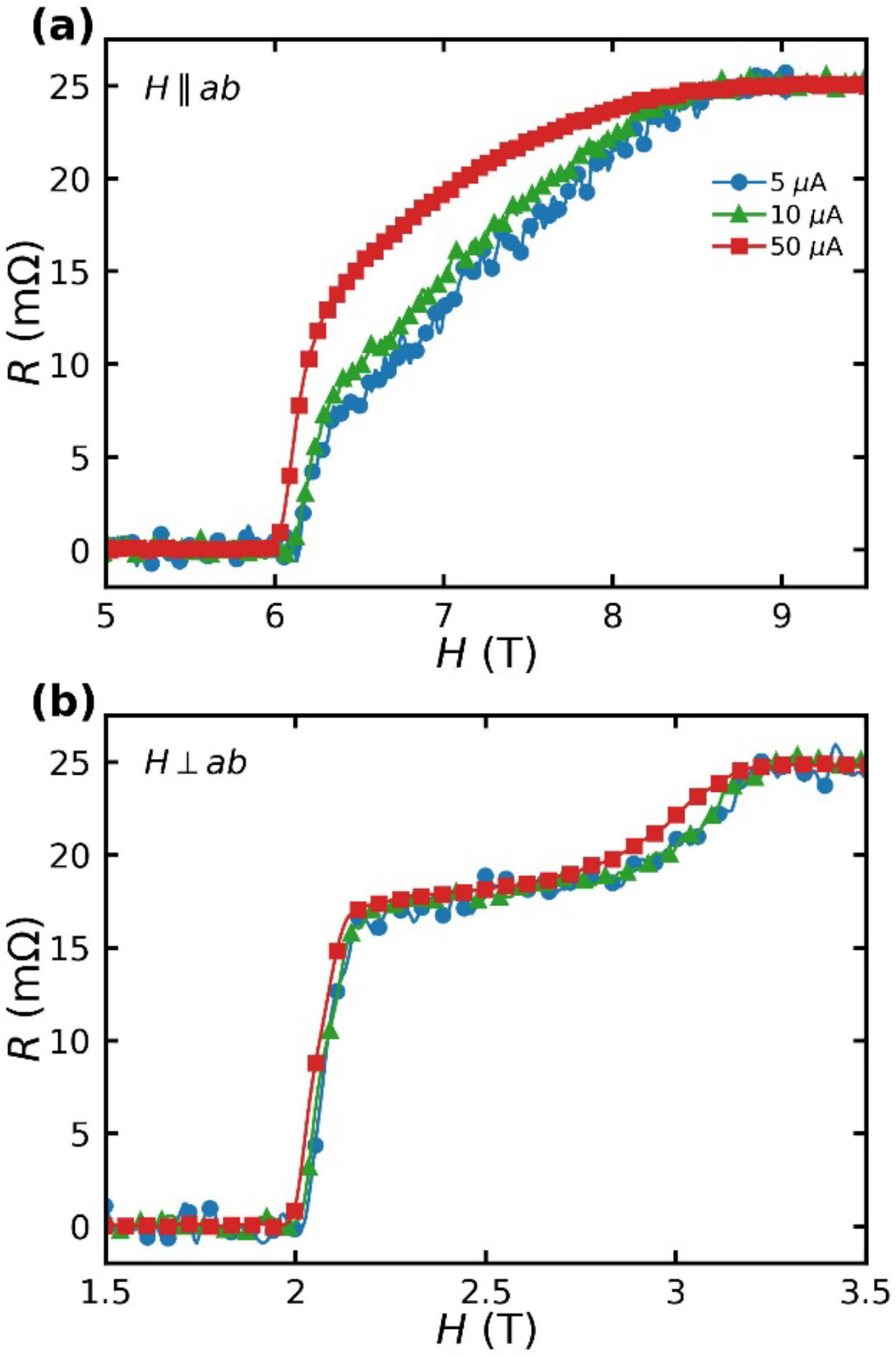

FIG 4. Magnetic-field dependence of the resistance $R$ measured with various excitation currents at an average temperature of 45 mK for (a) $H \parallel ab$ and (b) $H \perp ab$.

## IV. DISCUSSION

A notable feature of the present results is the broad transition for in-plane fields and the resultant criterion-dependent angular dependence of the critical field. The fields defined by the 10% $R_N$ criterion follow the anisotropic GL model, whereas those defined by the 65% and 90% $R_N$ criteria exhibit Tinkham-like angular dependence. The 10% criterion, located near the low-resistance side of the transition, is expected to predominantly reflect bulk superconductivity, whereas the higher-resistance criteria may be sensitive to a superconducting component that survives above the bulk $H_{c2}$. Similar criterion-dependent behavior has been reported in $MgB_2$ single crystals [19,20] and Nb films [21], where the high-resistance part of the transition was attributed to surface superconductivity. Surface superconductivity above the bulk $H_{c2}$ was theoretically predicted by Saint-James and de Gennes [22], and current-dependent broad resistive transitions associated with surface superconductivity have been experimentally observed in type-II superconductors [23], including layered superconductors such as $NbSe_2$ [24].

Several features of the present data support the existence of surface superconductivity in the present sample. First, the angular dependence obtained by extrapolating the lower-field transition is well described by the surface-superconductivity model (Fig. 3(b), bottom panel). Second, the intermediate- and high-resistance parts of the broad transition for in-plane fields exhibit a pronounced excitation-current dependence, whereas the low-resistance side is much less affected (Fig. 4(a)), consistent with a surface superconducting state having a limited current-carrying capacity. Third, the broad transition appears only for in-plane fields. Because the present material has a van der Waals layered structure, surface superconductivity is expected to develop preferentially on the large $ab$-plane surfaces, for which the parallel-field configuration is favorable. Taken together, these observations support the interpretation that the broad transition for in-plane fields, particularly its high-resistance part, is associated with surface superconductivity on the quasiperiodic $ab$-plane surfaces.

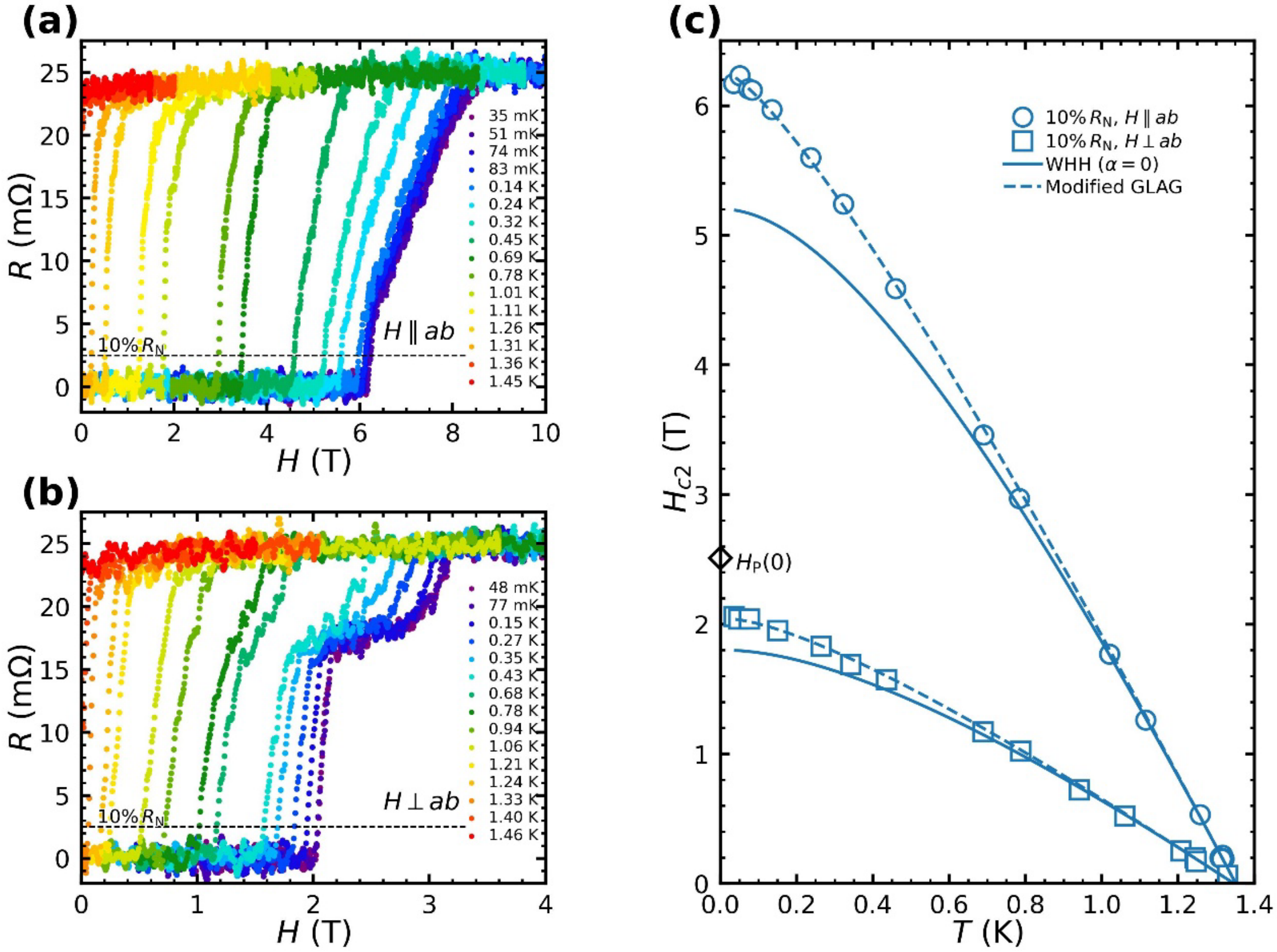

FIG 5. Magnetic-field dependence of the resistance $R$ at various temperatures for (a) in-plane fields and (b) out-of-plane fields. (c) Temperature dependence of the upper critical fields for in-plane and out-of-plane fields, determined using the 10% $R_N$ criterion. Open circles and open squares represent the data for in-plane and out-of-plane fields, respectively. Solid and dashed curves represent the orbital-limited WHH model ($\alpha = 0$) and the phenomenologically modified GLAG model, respectively. The diamond represents the Pauli limit $H_P(0)$.

If the critical fields determined using the 10% and 90% $R_N$ criteria for in-plane fields are regarded as characteristic fields of bulk and surface superconductivity, respectively, their ratio is approximately 1.38, smaller than the Saint-James–de Gennes value $H_{c3}/H_{c2}$ = 1.695 for an ideal isotropic semi-infinite superconductor [22]. However, resistively defined critical fields do not necessarily coincide with the thermodynamic $H_{c2}$ and $H_{c3}$. In particular, $H_{c3}$ should, in principle, correspond to the field at which $R(H)$ reaches $R_N$. Because $R(H)$ approaches $R_N$ only gradually, the critical field determined using the 90% $R_N$ criterion may substantially underestimate the actual $H_{c3}$. Furthermore, microscopic calculations have shown that $H_{c3}/H_{c2}$ can deviate from 1.695 depending on temperature and scattering [25]. Therefore, the difference from the ideal value does not necessarily contradict the surface-superconductivity interpretation.

In contrast to the pronounced current dependence of the broad transition for in-plane fields, the two-step transition for out-of-plane fields shows much weaker current dependence, as shown in Fig. 4(b), suggesting that the latter has a different origin from the former. The sharp superconducting transition observed in zero field (Fig. 2) indicates that the spatial variation in $T_c$ is small, making an inhomogeneous superconducting state an unlikely origin of the two-step transition. Furthermore, the absence of additional diffraction peaks in powder XRD and the lack of an additional superconducting transition in zero-field resistivity measurements do not support the presence of a secondary superconducting phase. Nevertheless, the presence of a trace superconducting impurity phase below the detection limit of the powder XRD measurements cannot be completely excluded. Further local structural and spectroscopic studies will be required to clarify the origin of the two-step transition for out-of-plane fields.

Another important feature of the present results is the anomalous temperature dependence of $H_{c2}$. Although the behavior near $T_c$ is well reproduced by the dirty-limit WHH theory with $\alpha = 0$, both $H_{c2}^{\parallel}$ and $H_{c2}^{\perp}$ deviate upward from the WHH predictions at lower temperatures. This behavior cannot be explained solely by strong spin–orbit scattering within the conventional single-band dirty-limit WHH framework, because spin–orbit scattering suppresses Pauli paramagnetic pair breaking but does not by itself produce an enhancement beyond the orbital-limited ($\alpha = 0$) behavior. A possible origin of the upward deviation is spatial inhomogeneity in the electronic diffusivity $D$. Carter *et al*. showed that spatial variations in $D$ on a length scale comparable to the superconducting coherence length can enhance $H_{c2}$ above the conventional GLAG prediction [18]. In a QC, quasiperiodic order gives rise to a variety of inequivalent local atomic environments, which may in turn produce spatial variations in electronic transport properties. To examine this possibility, we analyzed $H_{c2}(T)$ using the phenomenologically modified GLAG model proposed by Carter *et al.* [18], in which the electronic diffusivity $D$ is allowed to vary spatially. The reduced upper critical field is described by

$$\ln t = \psi\left(\frac{1}{2}\right) - \int \psi\left(\frac{1}{2} + \frac{\pi^2}{2}\frac{h}{t}y\right)Q(y)dy, \qquad (5)$$

where $t = T/T_c$, $h = \frac{H_{c2}}{T_c|dH_{c2}/dT|_{T_c}}$, $y = D/D_0$ is the local electronic diffusivity normalized by its characteristic value $D_0$, $\psi$ denotes the digamma function, and $Q(y)$ is the normalized distribution function. Following the form of the symmetric distribution employed by Carter *et al*., we used a generalized distribution given by

$$Q(y) = \frac{2}{\pi w^2}\sqrt{w^2 - (y-1)^2}, \qquad (6)$$

where $w$ is the dimensionless half-width of the electronic-diffusivity distribution and was treated as the only additional adjustable parameter. The fits yielded $w = 0.99$ for in-plane field and $w = 0.88$ for out-of-plane field. As shown in Fig. 5(c), the modified GLAG model provides a substantially improved description of the low-temperature $H_{c2}(T)$ for both field orientations. The agreement with the modified GLAG model suggests that spatial variations in the electronic diffusivity may contribute to the anomalous temperature dependence of $H_{c2}$ in the present QC.

The modified-GLAG analysis indicates that $H_{c2}$ is predominantly determined by orbital pair breaking for both field orientations. At the same time, $H_{c2}^{\parallel}$ exceeds the weak-coupling Pauli limit by a factor of approximately 2.5, indicating that conventional weak-coupling Pauli paramagnetic pair breaking does not determine $H_{c2}^{\parallel}$. This suppression of Pauli paramagnetic pair breaking is consistent with our previous $^{125}$Te-NMR measurements on the Cu-substituted QC $(Ta_{0.95}Cu_{0.05})_{1.6}Te$, which showed only a small decrease in the Knight shift below $T_c$, indicating only a small reduction in the spin susceptibility [26].

Similar large enhancements of $H_{c2}^{\parallel}$ beyond the Pauli limit have been reported for layered transition-metal dichalcogenides such as $NbSe_2$ and $TaS_2$, where they have been attributed to an Ising superconducting state protected by Zeeman-type spin–orbit coupling (SOC) [27,28]. Theoretical studies of superconductivity on a two-dimensional Penrose lattice have also demonstrated the possible coexistence of Rashba- and Ising-type SOCs in a quasiperiodic system [11]. In the Penrose lattice, Rashba-type SOC was found to suppress paramagnetic pair breaking for both in-plane and out-of-plane fields, whereas Ising-type SOC preferentially protects superconductivity against in-plane fields. Both types of SOC could thus contribute to the suppression of Pauli paramagnetic pair breaking in the present Ta–Te-based QC system. The present $H_{c2}$ measurements alone, however, do not allow us to identify the relevant SOC mechanism. Reducing orbital pair breaking, for example by reducing the number of layers through exfoliation or by weakening the interlayer coupling through intercalation, may provide a means of revealing the influence of SOC on the Pauli limiting field more directly.

## V. CONCLUSIONS

We have investigated the upper critical field of a large single grain of the Nb-substituted van der Waals layered QC $(Ta_{0.7}Nb_{0.3})_{1.6}Te$. A pronounced anisotropy of the upper critical field was observed. The angular dependence determined using the 10% $R_N$ criterion follows the anisotropic GL model, whereas those determined using the 65% and 90% $R_N$ criteria exhibit Tinkham-like behavior. When the 90% $R_N$ critical field near the out-of-plane direction are estimated by extrapolating the lower-field transition to account for the two-step transition, their angular dependence is well described by the surface-superconductivity model. Combined with the pronounced current dependence of the broad in-plane resistive transition, these observations support the presence of surface superconductivity on the quasiperiodic *ab*-plane surfaces. $H_{c2}^{\parallel}$ exceeds the weak-coupling Pauli limit by a factor of approximately 2.5, whereas $H_{c2}^{\perp}$ remains below it. Although the temperature dependence of $H_{c2}$ near $T_c$ is reproduced by the dirty-limit WHH theory with $\alpha$ =

0, both $H_{c2}^{\parallel}$ and $H_{c2}^{\perp}$ deviate upward from the WHH predictions at lower temperatures. A phenomenologically modified GLAG model incorporating a spatial distribution of the electronic diffusivity substantially improves the description of $H_{c2}(T)$, suggesting that spatial variations in electronic transport properties may contribute to its anomalous temperature dependence. These results provide new insight into the interplay between quasiperiodicity, layered structure, and superconductivity in van der Waals QCs.

## ACKNOWLEDGMENTS

The authors would like to thank T. Kobayashi and Y. Niimi for fruitful discussions. This study was supported by the JST-CREST program (grant no. JPMJCR22O3; Japan), JSPS KAKENHI Grant Number JP23K04355, and Iketani Science and Technology Foundation.

Supplemental Material

# Anisotropic upper critical field in the van der Waals superconducting quasicrystal $(Ta_{0.7}Nb_{0.3})_{1.6}Te$

Koki Kasai[1], Yuki Tokumoto[1,*], Taichi Terashima[2], Takako Konoike[2], and Keiichi Edagawa[1]

[1]*Institute of Industrial Science*, *The University of Tokyo*, *Tokyo 153-8505*, *Japan*
[2]*Research Center for Materials Nanoarchitectonics (MANA)*, *National Institute for Materials Science*, *Tsukuba 305-0003*, *Japan*

*Corresponding author: tokumoto@iis.u-tokyo.ac.jp

## 1. Calculation of the powder X-ray diffraction profile

In general, the structure of a crystalline approximant to a quasicrystal can be regarded as a periodic structure obtained by introducing phason strain into the corresponding quasicrystalline structure. Based on this relationship between a quasicrystal and its approximant, the powder X-ray diffraction (XRD) profile of the quasicrystal can be calculated from that of the approximant. The principle and procedure of this calculation are described in detail in our previous paper [1], where the powder XRD profile of the $Ta_{1.6}Te$ quasicrystal was calculated from that of the $Ta_{97}Te_{60}$ approximant. The results of that calculation are reproduced in Table S1.

The calculated profile shown in Fig. 1 of the main text was obtained from these data by additionally taking into account the effects of phason disorder and stacking disorder along the $c$-axis. Phason disorder is frequently present in quasicrystals. The structural quality of the $Ta_{1.6}Te$ quasicrystal is relatively low compared with that of Al-based well-ordered stable quasicrystals, indicating the presence of a substantial amount of phason disorder. Such phason disorder is expected to cause peak broadening that depends on the perpendicular-space component, $\mathbf{G}_\perp$, of the higher-dimensional reciprocal-lattice vector. In addition, stacking disorder along the $c$-axis gives rise to peak broadening that depends on the $c$-axis index $h_5$. Taking these effects into account, the powder XRD profile was calculated from the data in Table S1 as follows.

For a reflection indexed by $(h_1, \ldots, h_5)$, the perpendicular-space component $\mathbf{G}_\perp$ of the four-dimensional reciprocal-lattice vector is given by

$$\mathbf{G}_\perp = \sum_{i=1}^{4} h_i \mathbf{u}_i^* ,$$

where the perpendicular-space reciprocal basis vectors $\mathbf{u}_i^*$ ($i$=1,…,4) are given by

$$\mathbf{u}_i^* = -a^* \left( \cos\left(\frac{2\pi}{12} \cdot 5 \cdot (i-1)\right), \sin\left(\frac{2\pi}{12} \cdot 5 \cdot (i-1)\right) \right).$$

Here, $a^* = 0.6942\ \text{Å}^{-1}$ for the $Ta_{1.6}Te$ quasicrystal. The half width at half maximum (HWHM), $w$, of the diffraction peak was assumed to be

$$w = \{\alpha + \beta \cdot (G_\perp / a^*)\} \cdot (1 + \gamma \cdot |h_5|),$$

where $G_\perp = |\mathbf{G}_\perp|$, and $\alpha$, $\beta$, and $\gamma$ are real parameters. The calculated diffraction profile was then obtained by summing Lorentzian functions over all reflections, using the peak position $q_i$, integrated intensity $I_i$, and HWHM $w_i$ of the $i$-th reflection:

$$I(q) = \sum_i \frac{I_i}{\pi} \cdot \frac{w_i}{(q - q_i)^2 + w_i^2} .$$

Finally, $q$ was converted to the diffraction angle $2\theta$ using $q = \frac{4\pi \sin\theta}{\lambda}$, where $\lambda = 1.5405$ Å is the wavelength of the Cu $K\alpha$ radiation, to obtain the calculated powder XRD profile $I(2\theta)$. For the calculated profile shown in Fig. 1 of the main text, we used $\alpha = 0.0030$ Å$^{-1}$, $\beta = 0.019$ Å$^{-1}$, and $\gamma$=0.15. The calculated profile reproduces the experimental diffraction profile reasonably well.

Table S1. Indices, $q$-value, scattering angle $2\theta$, and relative intensity $I$ calculated for the $Ta_{1.6}Te$ quasicrystal. $q = \frac{4\pi \sin\theta}{\lambda}$ and $\lambda = 1.5405$ Å. Reproduced from Supplementary Table 3 of Ref. [1] under the Creative Commons Attribution 4.0 International License.

| $h_1$ | $h_2$ | $h_3$ | $h_4$ | $h_5$ | $q$ (Å$^{-1}$) | $2\theta$ (degree) | $I$ |
|---|---|---|---|---|---|---|---|
| 0 | 0 | 0 | 0 | 1 | 0.6047 | 8.50 | 100.00 |
| 1 | 0 | 0 | 0 | 0 | 0.6942 | 9.76 | 0.48 |
| 1 | -1 | 0 | 0 | 1 | 0.7034 | 9.89 | 5.47 |
| 1 | 0 | 0 | 0 | 1 | 0.9206 | 12.96 | 1.77 |
| 0 | 0 | 0 | 0 | 2 | 1.2094 | 17.05 | 0.49 |
| 1 | -1 | 0 | 0 | 2 | 1.2617 | 17.80 | 0.71 |
| 1 | 1 | 0 | 0 | 0 | 1.3411 | 18.93 | 1.81 |
| 1 | 0 | 0 | 0 | 2 | 1.3945 | 19.69 | 0.36 |
| 1 | 1 | 1 | 0 | 0 | 1.8966 | 26.89 | 3.87 |
| 2 | 0 | 0 | 1 | 2 | 1.9678 | 27.92 | 0.52 |
| 1 | 1 | 1 | 0 | 1 | 1.9906 | 28.25 | 7.70 |
| 2 | 1 | -1 | 0 | 2 | 2.0555 | 29.19 | 0.48 |
| 2 | 1 | 0 | 0 | 1 | 2.1082 | 29.96 | 0.42 |
| 2 | 1 | -1 | -1 | 2 | 2.2494 | 32.02 | 1.87 |
| 1 | 1 | 0 | 0 | 3 | 2.2560 | 32.11 | 0.56 |
| 2 | 0 | 0 | 0 | 3 | 2.2845 | 32.53 | 0.55 |
| 2 | 1 | 0 | 0 | 2 | 2.3541 | 33.55 | 1.44 |
| 2 | 0 | 0 | 1 | 3 | 2.3876 | 34.04 | 3.27 |
| 0 | 0 | 0 | 0 | 4 | 2.4189 | 34.50 | 0.24 |
| 1 | -1 | 0 | 0 | 4 | 2.4454 | 34.89 | 2.70 |
| 2 | 1 | -1 | 0 | 3 | 2.4604 | 35.11 | 6.05 |
| 1 | 0 | 0 | 0 | 4 | 2.5165 | 35.94 | 4.28 |
| 2 | 2 | -1 | 0 | 2 | 2.5506 | 36.44 | 4.67 |
| 1 | 2 | 1 | 0 | 0 | 2.5908 | 37.04 | 10.76 |
| 1 | 0 | 0 | 1 | 4 | 2.6105 | 37.33 | 4.43 |
| 2 | 2 | -1 | -1 | 2 | 2.6188 | 37.45 | 2.26 |
| 2 | 1 | -1 | -1 | 3 | 2.6245 | 37.54 | 4.81 |
| 2 | -1 | -1 | 0 | 4 | 2.6351 | 37.70 | 7.69 |
| 3 | 1 | -1 | 0 | 2 | 2.6433 | 37.82 | 4.73 |
| 1 | 2 | 1 | 0 | 1 | 2.6604 | 38.07 | 19.60 |
| 2 | 0 | -1 | 0 | 4 | 2.7012 | 38.68 | 3.08 |
| 2 | 1 | 0 | 0 | 3 | 2.7148 | 38.88 | 9.54 |
| 2 | 2 | 0 | 0 | 1 | 2.7495 | 39.40 | 0.25 |
| 1 | 1 | 0 | 0 | 4 | 2.7658 | 39.64 | 3.97 |
| 3 | 1 | 0 | 0 | 1 | 2.7729 | 39.75 | 0.40 |
| 2 | 0 | 0 | 0 | 4 | 2.7890 | 39.99 | 2.40 |
| 3 | 1 | -1 | -1 | 2 | 2.7968 | 40.10 | 3.80 |
| 2 | 2 | 0 | -1 | 2 | 2.8592 | 41.04 | 7.89 |
| 2 | 0 | 0 | 1 | 4 | 2.8741 | 41.26 | 3.84 |
| 2 | 2 | -1 | 0 | 3 | 2.8868 | 41.45 | 5.29 |
| 3 | 1 | -1 | -2 | 1 | 2.9195 | 41.94 | 0.48 |
| 2 | 1 | -1 | 0 | 4 | 2.9348 | 42.18 | 3.70 |
| 2 | 2 | 0 | 0 | 2 | 2.9422 | 42.29 | 0.59 |
| 2 | 2 | -1 | -1 | 3 | 2.9473 | 42.36 | 1.20 |
| 3 | 1 | 0 | 0 | 2 | 2.9641 | 42.62 | 1.49 |
| 3 | 1 | -1 | 0 | 3 | 2.9691 | 42.69 | 2.99 |
| 3 | 2 | -1 | 0 | 1 | 3.0009 | 43.17 | 0.43 |
| 0 | 0 | 0 | 0 | 5 | 3.0236 | 43.52 | 0.20 |
| 1 | -1 | 0 | 0 | 5 | 3.0449 | 43.84 | 0.72 |
| 3 | 2 | -1 | -1 | 1 | 3.0591 | 44.05 | 0.23 |
| 2 | 1 | -1 | -1 | 4 | 3.0737 | 44.28 | 0.92 |
| 3 | 1 | -1 | -2 | 2 | 3.1017 | 44.70 | 0.45 |
| 1 | 0 | 0 | 0 | 5 | 3.1023 | 44.71 | 3.07 |
| 3 | 1 | -1 | -1 | 3 | 3.1065 | 44.77 | 0.57 |
| 2 | 1 | 0 | 0 | 4 | 3.1512 | 45.45 | 1.12 |
| 2 | 2 | 0 | -1 | 3 | 3.1628 | 45.63 | 0.25 |
| 1 | 0 | 0 | 1 | 5 | 3.1790 | 45.88 | 0.94 |
| 2 | -1 | -1 | 0 | 5 | 3.1992 | 46.18 | 2.07 |
| 3 | 2 | -1 | -2 | 0 | 3.2108 | 46.36 | 0.28 |
| 2 | 0 | -1 | 0 | 5 | 3.2539 | 47.02 | 0.78 |
| 3 | 2 | -1 | -2 | 1 | 3.2672 | 47.22 | 0.21 |
| 3 | 2 | 0 | -1 | 0 | 3.2849 | 47.50 | 0.71 |
| 2 | 2 | -1 | 0 | 4 | 3.3005 | 47.74 | 0.61 |
| 1 | 1 | 0 | 0 | 5 | 3.3077 | 47.85 | 0.49 |
| 3 | 2 | 0 | -1 | 1 | 3.3401 | 48.35 | 0.90 |

**2. Single-crystalline sample**

Figure S1 shows optical microscope and scanning electron microscope (SEM) images of the single-crystalline grains obtained after heat treatment. In addition to the sintered pellet of approximately 10 mm in diameter, relatively large plate-like grains were found to have grown on the inner wall of the Mo crucible, as shown in Fig. S1(a). Figure S1(b) shows a cross-sectional optical microscope image of one of these grains. Figure S1(c) shows an SEM image of a cleaved surface of a representative grain. SEM-EDS analysis yielded an Nb/(Ta+Nb) atomic ratio of approximately 26%.

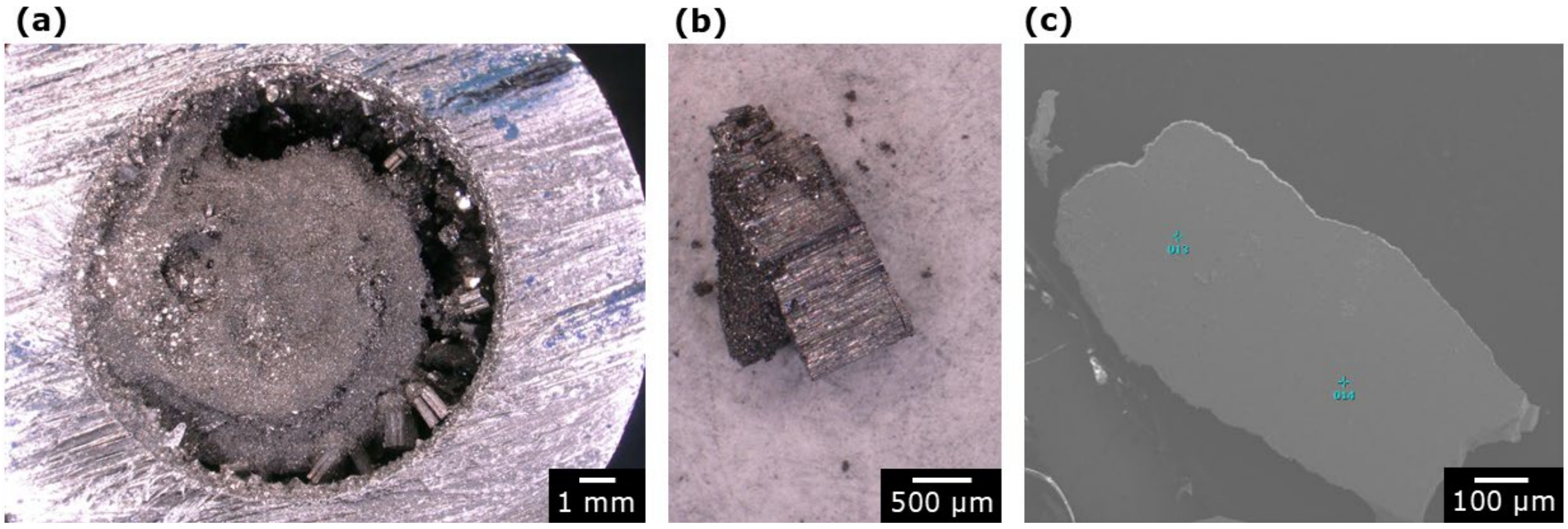

Figure S1. (a) Optical microscope image of the sample in the Mo crucible after heat treatment. (b) Cross-sectional optical microscope image of a large single-crystalline grain. (c) SEM image of a cleaved surface of the single-crystalline grain.